# Voltage-controlled skyrmion-based artificial synapse in a synthetic antiferromagnet

Ziyang Yu[1], Maokang Shen[2,#], Zhongming Zeng[3], Shiheng Liang[4], Yong Liu[1], Ming Chen[1], Zhenhua Zhang[1], Zhihong Lu[5], Yue Zhang[2,*], Rui Xiong[1,*]

1. Key Laboratory of Artificial Micro- and Nano-structures of Ministry of Education, School of Physics and Technology, Wuhan University, Wuhan, 430072, P. R. China
2. School of Optical and Electronic Information, Huazhong University of Science and Technology, Wuhan, 430074, P. R. China
3. Key Laboratory of Multifunctional Nanomaterials and Smart Systems, Suzhou Institute of Nano-Tech and Nano-Bionics, Chinese Academy of Sciences, Suzhou, Jiangsu, 215123, P. R. China
4. Department of Physics, Hubei University, Wuhan 430062, P. R. China.
5. The State Key Laboratory of Refractories and Metallurgy, School of Materials and Metallurgy,Wuhan University of Science and Technology, Wuhan, 430081, P. R. China

*The corresponding author: Yue Zhang (E-mail: yue-zhang@hust.edu.cn) and Rui Xiong (E-mail: xiongrui@whu.edu.cn)

#The author who has the same contribution to Ziyang Yu

**Abstract**

Spintronics exhibits significant potential in neuromorphic computing system with high speed, high integration density, and low dissipation. In this letter, we propose an ultralow-dissipation spintronic memristor composed of a synthetic antiferromagnet (SAF) and a piezoelectric substrate. Skyrmions/skyrmion bubbles can be generated in the upper layer of SAF with weak anisotropy energy ($E_a$). With a weak electric field on the heterostructure, the interlayer antiferromagnetic coupling can be manipulated, giving rise to a continuous transition between a large skyrmion bubble and a small skyrmion. This thus induces the variation of the resistance of a magnetic tunneling junction. The synapse based on this principle may manipulate the weight in a wide range at a cost of a very low energy consumption of 0.3 fJ. These results pave a way to ultralow power neuromorphic computing applications.

Artificial neural network (ANN), which is based on a complicated connection of artificial neurons and synapse to mimic the neural network in the brain, has attracted extensive interest due to its powerful intelligence on information processing. However, the energy consumption of the ANN system developed on the traditional microelectronic element, such as Complementary Metal Oxide Semiconductor (CMOS) is much larger than that of the brain. Thereover, several functional devices are candidates for the implementation of ANN, such as memristors [1], phase change memory [2, 3], ferroelectric tunneling junction [4], and spintronic devices [5-7]. Among them, spintronic devices with the advantages of high speed, high integration degree, and low energy consumption unveil the great potential for developing novel ANN systems, while extensive research about ANN applications based on spintronic devices such as spin-torque oscillators (STOs) [5, 6], spintronic memristors and neurons [8-13] have been proposed.

Research on spintronic memristors, which used to construct synapses in a neural network, is one of the major research directions. A typical realization is based on the current-induced domain wall motion in the free layer of an MTJ, in which the resistance can be tuned by adjusting the percentage of areas of two different domains separated by the domain wall, to mimic the plasticity of synapses [8-10]. Therefrom, spintronic memristor based on the current-induced motion of skyrmions is proposed, showing the accessibility to imitate the potentiation and depression of synapses [14, 15]. However, the current-induced dissipation may hinder the applications of spintronic memristors. Recently, research on the manipulation of magnetization by electric field leads to a low power method to change the magnetic structure in a multiferroic heterostructure composed by ferroelectric and ferromagnetic materials [16-18]. In general, the magnetic structure is modified via an electric-field-induced variation of magnetic anisotropy energy [19, 20]. Very recently, Wang et al. report that the Ruderman-Kittel-Kasuya-Yosida (RKKY) exchange coupling strength in a synthetic antiferromagnet (SAF) on a piezoelectric substrate can be also effectively tuned by electric field [21].

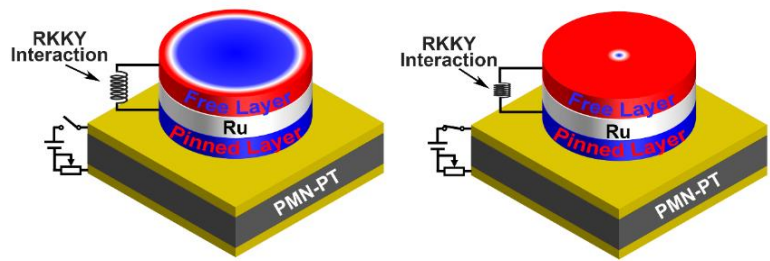

**Fig. 1. Schematic of the transition between a skyrmion bubble and a skyrmion based on manipulation of interlayer RKKY exchange interaction under an external voltage**

In this work, we report that the electric-field-induced variation of the RKKY exchange coupling strength can give rise to a transition between a big skyrmion bubble and a small skyrmion with ultralow dissipation. This results in a wide range variation of resistance of MTJ, which provides a way to build up spintronic synapse in an energy-efficiency way.

**Simulation methodology.** Micromagnetic simulation is carried out using Object-Oriented Micromagnetic-Framework (OOMMF) software with the code of interfacial Dzyaloshinskii-Moriya Interaction (DMI) [22]. The model is a circular SAF multilayer composed by ferromagnetic (FM) Pt/Co layers or ultrathin CoFeB layers with perpendicular magnetic anisotropy (PMA). They are two representative compositions for numerical investigation about skyrmions. The two FM layers in SAF have distinct magnetic anisotropy energy (Fig. 1). The magnetic moments in the lower FM layer are parallel aligned due to its strong anisotropy energy, while a skyrmion bubble/skyrmion can be generated in the upper layer due to its weaker anisotropy energy.

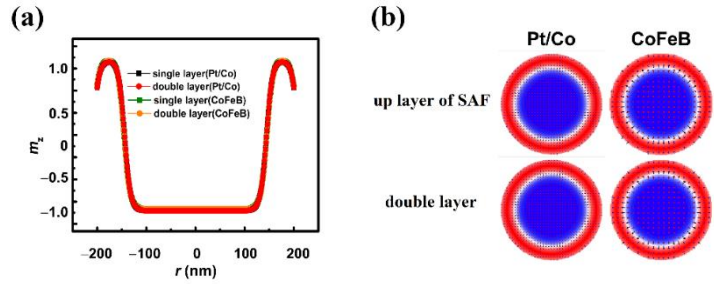

**Fig. 2. Structure of a skyrmion bubble in a single FM circular plate and that in a SAF with no interlayer exchange coupling: (a) The $m_z$ as a function of polar coordinate $r$ and (b) the images for the skyrmion bubbles in the single layer and the upper layer of SAF.**

The parameters for simulation are as following: the radius ($R$) of the plate is between 50 and 200 nm. The thicknesses of the lower, the upper, and the interlayer are all 0.4 nm, respectively. The cell size is 1 nm × 1nm × 0.4 nm. The saturation magnetization ($M_S$) of Pt/Co and CoFeB is $5.8 \times 10^5$ A/m and $1.1 \times 10^6$ A/m, respectively. The damping coefficient $\alpha$ of Pt/Co and CoFeB is 0.3 and 0.03, respectively [23, 24]. The exchange stiffness constant ($A$) is fixed as $1.5 \times 10^{-11}$ J/m. The magnetic anisotropy constant of the lower FM ($K_L$) of Pt/Co and CoFeB is fixed at $5 \times 10^6$ J/m$^3$ and $1 \times 10^7$ J/m$^3$, respectively. The magnetic anisotropy constant for the upper layer ($K_U$) is between $1 \times 10^5$ J/m$^3$ and $6 \times 10^5$ J/m$^3$, and the DMI constant ($D$) varies between 2 mJ/m$^2$ and 4 mJ/m$^2$. In experiments, the anisotropy constant and DMI constant can be manipulated by tuning the thickness of the heavy metal layer or the FM layer. The interlayer antiferromagnetic coupling energy density ($J_{ex}$) between two FM layers is between $-1.0 \times 10^{-6}$ J/m² and $-1.0 \times 10^{-4}$ J/m². For an SAF multilayer (CoFeB/Ru/CoFeB) on a piezoelectric substrate, the $J_{ex}$ can be manipulated by an external electric field [21]. Here, the electric-field-induced changing of magnetic anisotropy energy is neglected, of which the validity will be discussed below.

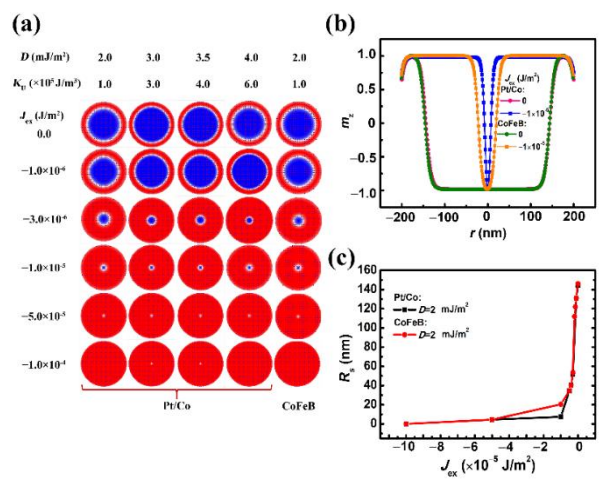

**Fig. 3.** (a). Snapshots for the transition between a skyrmion bubble and a skyrmion under different $D$ and $K_U$ for Pt/Co and CoFeB; (b). variation of $m_z$ with polar coordinate $r$ under different $J_{ex}$ ($D = 2$ mJ/m², and $K_U = 1 \times 10^5$ J/m³) for Pt/Co and CoFeB; (c). $R_s$ of skyrmion bubble/skyrmion as a function of $J_{ex}$ for Pt/Co and CoFeB.

**Transition between a big skyrmion bubble and a small skyrmion.** Initially, the single circular FM plate with an $R$ of 200 nm was divided into two concentric circles. The magnetic moments of the inner one with a radius of 170 nm were set to be along the $+z$ axis, while the remaining were along the $-z$ direction. After a relaxation for several nanoseconds, a typical skyrmion bubble with the radius ($R_s$) of 165 nm was generated. The moments inside are parallel to $+z$ direction, and a sharp transition of moment orientation from $+z$ to $-z$ direction happens across a Néel-typed DW at the boundary of the bubble. The canting of moments near the edge of the plate is due to the DMI-relevant edge effect [22]. As a comparison, we also performed the same simulation of the upper layer (The $K_U$ of Pt/Co and CoFeB are $4 \times 10^5$ J/m³ and $1 \times 10^5$ J/m³, respectively, and $D = 3.5$ mJ/m².) of SAF with $J_{ex} = 0$. The structure of the skyrmion bubble in this SAF is almost the same as its counterpart in the single FM layer (Fig. 2). Therefore, the stray field from the ultrathin lower layer of SAF had little impact on the structure of the skyrmion bubble. On the other hand, the size of the skyrmion bubble for Pt/Co and CoFeB is almost the identical.

When $J_{ex}$ is considered, the big skyrmion bubble is converted into a small skyrmion while increasing $J_{ex}$ (Fig. 3(a)), yet the influence of $D$ and $K_U$ in a wide range ($D = 2 \sim 4$ mJ/m$^2$, and $K_U = 1 \times 10^5 \sim 6 \times 10^5$ J/m$^3$) is negligible. When the $D$ is higher than 4 mJ/m$^2$, skyrmion bubble will be destabilized and transform to vortex [22]. It is also noticed that the reduction of the $R_s$ of skyrmion bubble occurs mainly under the weak interlayer exchange coupling when the $J_{ex}$ is between $-1.0 \times 10^{-6}$ J/m$^2$ and $-3.0 \times 10^{-6}$ J/m$^2$. With a stronger interlayer coupling, the $R_s$ is greatly reduced to < 30 nm. When the interlayer exchange coupling is very weak ($J_{ex} < -1.0 \times 10^{-5}$ J/m$^2$ or $J_{ex} > -5.0 \times 10^{-5}$ J/m$^2$), the $R_s$ of skyrmion bubble for Pt/Co and CoFeB is very close. In between, the size of CoFeB is a little larger than that of Pt/Co. In application, effective manipulation of $R_s$ by $J_{ex}$ indicates the possibility for tuning device behavior under the low electric field. This is good for reducing dissipation, and it will be discussed in detail below.

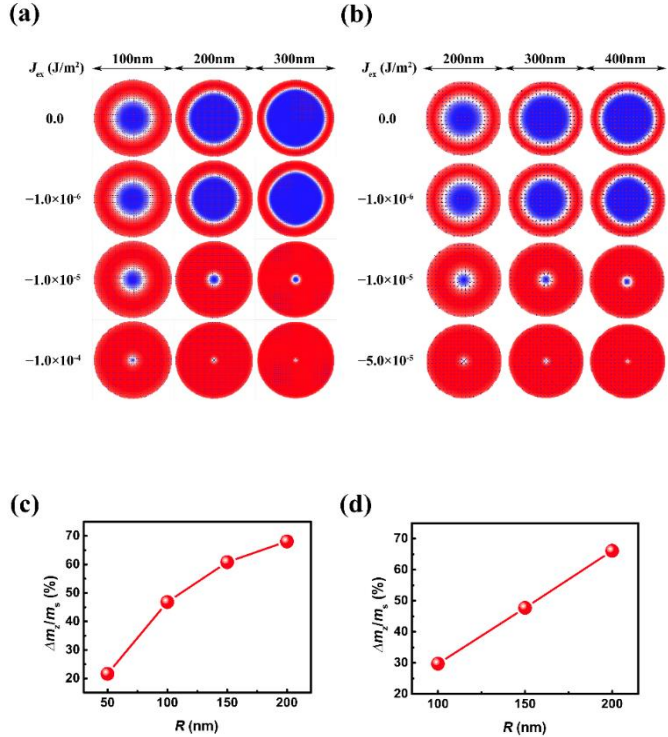

**Fig. 4.** Snapshot for the transition between skyrmion bubble and skyrmion in the circular plates with different $R$ for (a) Pt/Co ($D = 4$ mJ/m$^2$ and $K_L = 6 \times 10^5$ J/m$^3$) and (b) CoFeB ($D = 2$ mJ/m$^2$ and $K_L = 1 \times 10^5$ J/m$^3$); Relative variation of net magnetization of the upper FM layer with $R$ in the process of skyrmion bubbles-skyrmions transition for (c) Pt/Co and (d) CoFeB.

In addition to dissipation, the device size is another important factor for applications. The transition between the skyrmion bubble and skyrmion in the circular plates with different $R$ (50 nm ~ 200 nm) was also simulated (Fig. 4). In all sizes, the $R_s$ of skyrmion bubble decreases mainly when $J_{ex}$ is between $-1.0 \times 10^{-6}$ J/m$^2$ and $-1.0 \times 10^{-5}$ J/m$^2$, which is not depended on plate size. To characterize the resistance variation of an MTJ, we consider the relative variation of the area of skyrmion bubble with $J_{ex}$ increasing from 0 to $-1.0 \times 10^{-4}$ J/m$^2$ ($-5.0 \times 10^{-5}$ J/m$^2$) for Pt/Co (CoFeB). This can be reflected from the relative change of net magnetization in the upper layer, which is defined as $\Delta M_z/M_S$, where $\Delta M_z$ is the change of magnetization in the transition, and $M_S$ is the saturation magnetization. This $\Delta M_z/M_S$ is reduced with decreasing size of the plate. However, as to Pt/Co, when the $R$ of the plate is larger than 100 nm, it is still larger than 50 %, showing the possibility of an effective manipulation of resistance. For CoFeB, to obtain the $\Delta M_z/M_S$ that is larger than 50 %, the $R$ needs to increase to > 150 nm.

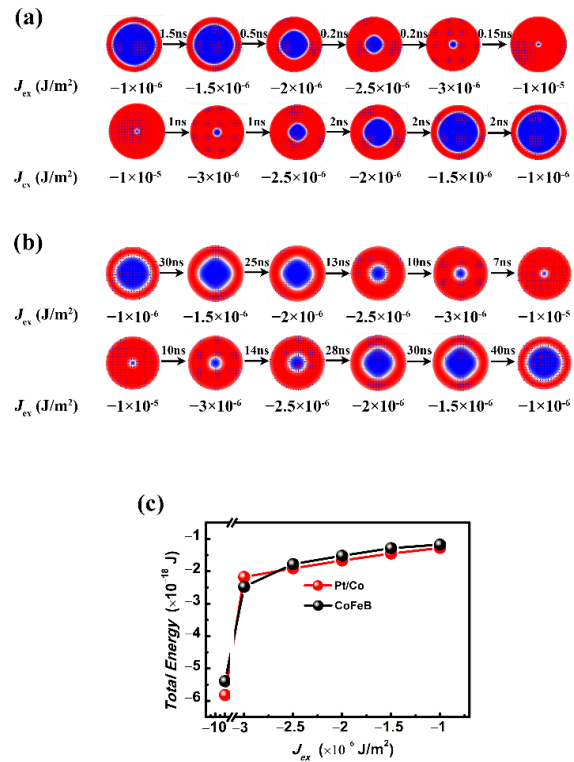

**Fig. 5.** Relaxation time for $R_s$ variation due to the change of $J_{ex}$ for (a) Pt/Co ($D$ = 4 mJ/m$^2$ and $K_L$ = 6 × 10$^5$ J/m$^3$) and (b) CoFeB ($D$ = 2 mJ/m$^2$ and $K_L$ = 1 × 10$^5$ J/m$^3$); (c). variation of *Total Energy* with $J_{ex}$ for Pt/Co and CoFeB

To accelerate the computing speed, the relaxation time ($\tau$) for stabilizing a skyrmion bubble/skyrmion due to the variation of $J_{ex}$ should be considered (Fig. 5). Initially, a stable skyrmion bubble with $J_{ex} = -1.0 \times 10^{-6}$ J/m² was generated in a circular plate with $R = 200$ nm. Then, the $\tau$ versus $J_{ex}$ ranging from $-1.0 \times 10^{-6}$ J/m² to $-1.0 \times 10^{-5}$ J/m² was recorded. The $\tau$ for Pt/Co and CoFeB is quite different. For Pt/Co, the $\tau$ in this processing is 2 ns or shorter, which means the device has a high response speed under an ultrashort voltage pulse. Nevertheless, the $\tau$ of CoFeB is longer than 10 ns, but the $\tau$ of tens of ns is still very short for a memristor. On the other hand, the relaxation time for increasing $R_s$ is longer than that for reducing $R_s$, which is originating from the variation of total free energy with $J_{ex}$ (Fig. 5(b)). The enhancement of RKKY interaction results in a decrease in energy, which corresponds to a faster relaxation process.

**Theoretical analysis.** In theory, every stable skyrmion bubble/skyrmion with different $R_s$ corresponds to the state with the minimum free energy of the SAF. For a big skyrmion bubble, the free energy densities including exchange ($E_{exch}$), uniaxial magnetic anisotropy ($E_{anis}$), DMI ($E_{DMI}$), demagnetization ($E_d$), RKKY ($E_{RKKY}$), and edge energy ($E_{edge}$) are expressed as follows [25]:

$$E_{exch} = 4\pi rt\sqrt{AK_{eff}} + \frac{8\pi At}{\frac{2r}{\pi\Delta}+1}$$

with $K_{eff} = K_L - \frac{1}{2}\mu_0 M_S^2$,

and $\pi\Delta = \sqrt{\frac{A}{K_{eff}}}$ (1);

$$E_{anis} = 4\pi rt\sqrt{AK_{eff}} \quad (2);$$

$$E_{DMI} = -2\pi^2 Drt \quad (3);$$

$$E_d = -2\pi\mu_0 M_s^2 I(d)rt^2$$

with $I(d) = -\frac{2}{3\pi}[d^2 + (1-d^2)\frac{E(u^2)}{u} - \frac{K(u^2)}{u}], d = \frac{2r}{t}, u^2 = \frac{d^2}{1+d^2}$ (4);

$$E_{RKKY} = 2\pi Jr^2 \quad (5);$$

$$E_{edge} = kr^2 t \quad (6).$$

Here, $t$ is the thickness of the upper layer with skyrmion bubble/skyrmion; $r$ is the distance from the center of the plate; $K(u)$ and $E(u)$ are the complex elliptic integrals of the first and second kind for determining the demagnetization of a bubble system. The $K_{eff}$ here is the effective magnetic anisotropy constant, which mainly depends on the $K_L$, while the $K_U$ is very small comparing to the strong anisotropy energy of the lower layer and the interlayer exchange coupling. The edge energy originates from the interaction between the DMI-induced canting moments at the edge of the plate and the moments at the boundary of a skyrmion bubble. The edge energy is assumed to satisfy a quadratic function with $r$ so that it increases drastically when the boundary of the skyrmion bubble moves close to the plate edge. The $k$ is the coefficient of edge energy.

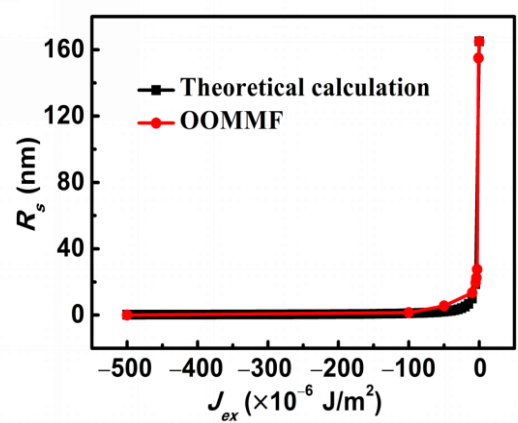

**Fig. 6. Comparison for $R_s$ variation of skyrmion bubble with $J_{ex}$ by theoretical analysis and simulation by OOMMF.**

The total free energy ($E$) is the summation from (1) to (6), and the $R_s$ for a stable skyrmion bubble/skyrmion is obtained by resolving the equation with $r$ as a variable:

$$\frac{dE}{dr} = 0 \qquad (7).$$

$k$ is firstly determined from the simulation result for the size of the skyrmion bubble with zero $J_{ex}$. Afterwards, based on all the parameters in simulation ($A=1.5 \times 10^{-11}$ J/m; $t = 0.4$ nm; $D = 4$ mJ/m$^2$; $K_b = 5 \times 10^6$ J/m$^3$; $M_S=5.8 \times 10^5$ A/m; $J_{ex} = 0 \sim -5 \times 10^{-4}$ J/m$^2$), the $R_s$ at different $J_{ex}$ is calculated using Eq. (7). As shown in Fig. 6, the solution of (7) is comparable to the simulation result. The small difference may be attributed to the error in calculating the edge energy.

**The spin memristor based on voltage-induced transition between a skyrmion bubble and a**

**skyrmion.** Based on the result of the simulation, we propose spintronic memristor based on a CoFeB MTJ deposited on a PMN-PT piezoelectric substrate. The free layer of MTJ is the upper layer of SAF. The parameters are the same as that in Fig. 5, and the $J_{ex}$ is tuned between $-1.0 \times 10^{-6}$ J/m$^2$ and $-1.0 \times 10^{-5}$ J/m$^2$ (Fig. 7). The maximum tunneling magnetoresistance (TMR) of MTJ is assumed to be 200 %. In the upper layer of SAF, the percentage of the area of skyrmion bubble/skyrmion with respect to that of the plate is assumed to be $x$, and the relationship between TMR and $x$ is assumed to satisfy a linear function [26]. The TMR increases from around 60 % to close to be close to 200 % with $J_{ex}$ increasing from $-1.0 \times 10^{-6}$ J/m$^2$ to $-5.0 \times 10^{-5}$ J/m$^2$.

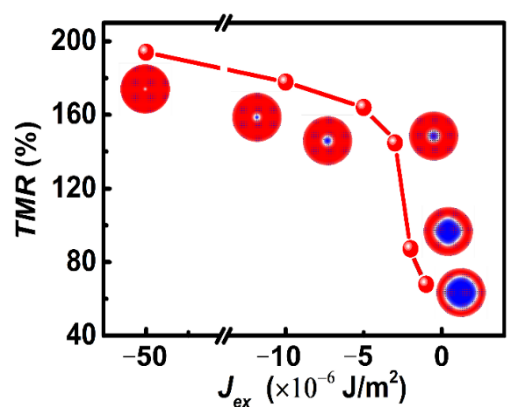

**Fig. 7. Variation of resistance in the transition between skyrmion bubble and skyrmion with change of $J_{ex}$ for CoFeB ($D = 2$ mJ/m$^2$ and $K_L = 1 \times 10^5$ J/m$^3$).**

In experiments, the $J_{ex}$ may be effectively tuned by the electric field. Based on the experimental results of the manipulation of RKKY exchange coupling in CoFeB/Ru/CoFeB SAF by electric field [21], we estimated the electric field strength and the dissipation for tuning resistance shown in Fig. 7. When the electric field is absence, $J_{ex} = -1.0 \times 10^{-5}$ J/m$^2$ by choosing a suitable thickness of the nonmagnetic layer. According to the slope of the variation of $J_{ex}$ with respect to electric field strength ($2.2 \times 10^{-5}$ (J/m$^2$)/(kV/cm)) [21], the electric field strength for reducing $J_{ex}$ from $-1.0 \times 10^{-5}$ J/m$^2$ to $-1.0 \times 10^{-6}$ J/m$^2$ should be around 0.4 kV/cm. The weak electric field cannot change the magnetic anisotropy constant but strong enough to manipulate $J_{ex}$ so that a small skyrmion can be converted into a large skyrmion bubble. This voltage-controlled multi-state behavior is analogous to the information transmission characteristics of biological synapses.

The operating dissipation of the skyrmion-based spintronic memristor can be estimated. We consider a circular PMN-PT substrate with the $R$ = 200 nm and the thickness $t$ = 100 μm. Based on the relative dielectric constant of PMN-PT (3000) [27] and $E$ = 0.4 kV/cm, the dissipation for the transition between a skyrmion bubble and a skyrmion is estimated to be smaller than $3 \times 10^{-16}$ J (0.3 fJ), which is much less than the magnitude of dissipation from the same transition triggered by magnetic field or current.

**Summary and outlook.** In conclusion, the transition between a big skyrmion bubble and a small skyrmion based on the varying interlayer RKKY antiferromagnetic coupling was studied. This transition mainly happens in a small range of weak RKKY exchange coupling and is not sensitive to the parameters such as DMI and magnetic anisotropy. Since the RKKY exchange coupling of a CoFeB SAF can be modulated by the multiferroic behavior, the highly efficient voltage control of RKKY exchange coupling is proved, giving rise to build up skyrmion-based spintronic memristor with ultralow energy consumption. These results unveil the potential for developing novel artificial synapse device with small size, high computing speed, and ultra-low dissipation. Therefore, further experimental investigation is very expected. On the other hand, when compared to CoFeB, the artificial synapse based on Pt/Co seems to have advantages in high computing speed. However, the experimental work about voltage-induced manipulation of RKKY in Pt/Co SAF has not been reported. This also deserves further investigation.

**Acknowledgements.** The authors would like to acknowledge the financial support from the National Natural Science Foundation of China (Nos. 11574096, 11774270, 11474225, 61674062) and Huazhong University of Science and Technology (No. 2017KFYXJJ037).